\documentclass[%
reprint,
superscriptaddress,
amsthm,amsmath,amssymb,
aps,
]{revtex4-1}

\usepackage{graphicx}
\usepackage{dcolumn}
\usepackage{bm}
\usepackage{physics}
\usepackage{mathrsfs}
\usepackage{amssymb}
\usepackage{subfigure}  
\usepackage{booktabs}
\usepackage{array}
\usepackage{amsmath}
\usepackage[dvipsnames]{xcolor}

\begin{document}
	
	\title{Exceptional-Point-Induced Sensitivity-Robustness Phase Transition in Quantum Interference }

	\author{Xing Lin}
	\affiliation{New Cornerstone Science Laboratory, Department of Physics, The University of Hong Kong, Hong Kong, 999077, China.}
	\author{Shuang Zhang}
	\email{To whom correspondence should be addressed \\ shuzhang@hku.hk}
	\affiliation{New Cornerstone Science Laboratory, Department of Physics, The University of Hong Kong, Hong Kong, 999077, China.}
	\affiliation{Department of Electrical \& Electronic Engineering, The University of Hong Kong, Hong Kong 999077, China.}
	\affiliation{State Key Laboratory of Optical Quantum Materials, University of Hong Kong, Hong Kong 999077, China.}
	\affiliation{Materials Innovation Institute for Life Sciences and Energy (MILES), HKU-SIRI, Shenzhen 518000, China.}
	\affiliation{Quantum Science Center of Guangdong-Hong Kong-Macao Great Bay Area, 3 Binlang Road, Shenzhen, China}

	\begin{abstract}		
		Quantum interference underpins many quantum information protocols but is typically studied in lossless Hermitian systems. Here, we reveal an exceptional-point-induced phase transition in two-photon Hong-Ou-Mandel interference within a lossy coupled-waveguide system. In the PT-symmetric phase, interference is ultrasensitive to coupling strength, yielding sharp bunching-antibunching switches. In the PT-broken phase, it becomes robust—oscillation-free and propagation-independent—with coincidence probability stably tunable via coupling. These regimes enable enhanced quantum sensing and reliable two-photon control for robust quantum information processing.
	\end{abstract}

\pacs{Valid PACS appear here}
\maketitle

Non-Hermitian physics has rapidly emerged as a vibrant and interdisciplinary field, revealing a wealth of peculiar phenomena across optical \cite{ozdemir2019parity,feng2017non,miri2019exceptional,el2018non}, acoustic \cite{tang2020exceptional}, electronic \cite{assawaworrarit2017robust,zhao2024exceptional}, and mechanical \cite{xu2016topological,wu2023chip} systems. Unlike conventional Hermitian quantum mechanics, which presumes energy conservation and closed systems, non-Hermitian systems incorporate gain and loss, leading to complex eigenvalues and fundamentally altered dynamics. Among their most striking features are exceptional points (EPs): singularities in parameter space where two or more eigenvalues and their corresponding eigenstates coalesce. At EPs, parity–time (PT) symmetry can be spontaneously broken, driving a phase transition from real to complex spectra and fundamentally changing the system’s behavior. These EP-induced transitions, which have no Hermitian counterpart, underpin a host of novel effects in classical regimes, such as non-orthogonal eigenmodes \cite{ruter2010observation,ching1998quasinormal}, topological chirality \cite{zeuner2015observation,weimann2017topologically}, unidirectional invisibility \cite{doppler2016dynamically}, robust lasing \cite{hodaei2014parity,feng2014single}, and enhanced sensing \cite{hodaei2017enhanced,chen2017exceptional}. Most research on non-Hermitian systems has been limited to classical and single-particle regimes \cite{lu2025quantum}, while investigations into genuinely quantum effects, including features such as photon indistinguishability and quantum correlations, remain scarce, making this an especially intriguing and promising field.

Quantum correlations and interference—most prominently Hong–Ou–Mandel (HOM) interference, where indistinguishable photons bunch to form maximally path-entangled states \cite{hong1987measurement}—lie at the heart of quantum information processing \cite{bouwmeester1997experimental}, communication \cite{lo2012measurement}, and computation \cite{hangleiter2023computational}. Traditionally these effects have been studied in Hermitian systems, where loss is considered purely detrimental. Surprisingly, controlled loss can dramatically reshape quantum statistics, even converting photon bunching into antibunching \cite{vest2017anti,hong2024loss,li2021non}. Recent experiments have realized two-photon HOM interference in lossy PT-symmetric photonic waveguides \cite{klauck2019observation,ehrhardt2022observation,klauck2025crossing,wolterink2023order} and performed quantum simulations of non-Hermitian Hamiltonians \cite{maraviglia2022photonic}, including gradual interference change cross the EP \cite{klauck2025crossing}; yet, these studies have largely remained within or near the PT-symmetric regime.

In this Letter, we reveal an EP-induced phase transition in two-photon quantum interference within a lossy coupled waveguide system—from ultrasensitive oscillations to robust, propagation-independent behavior—that has not been previously uncovered. In the PT-symmetric phase, the interference exhibits extreme sensitivity to coupling strength, producing rapid and counterintuitive switches between photon bunching and antibunching via sharp intra-group doublets. In the PT-symmetry-broken phase, by contrast, imaginary eigenvalues suppress all oscillations, yielding coincidence probabilities that are highly insensitive to both coupling strength and propagation distance, while remaining continuously tunable by the coupling parameter. Unlike Hermitian systems, which display regular periodic interference, the presence of the EP qualitatively reshapes the two-photon evolution on either side of the transition. These contrasting regimes not only enable orders-of-magnitude enhancement of quantum sensing sensitivity but also provide stable, tunable two-photon interference—capabilities that promise significant advantages for robust quantum information processing and high-precision metrology.
\\
\hfill

\begin{figure*}[htbp]
	\centering
	\includegraphics[width=7in]{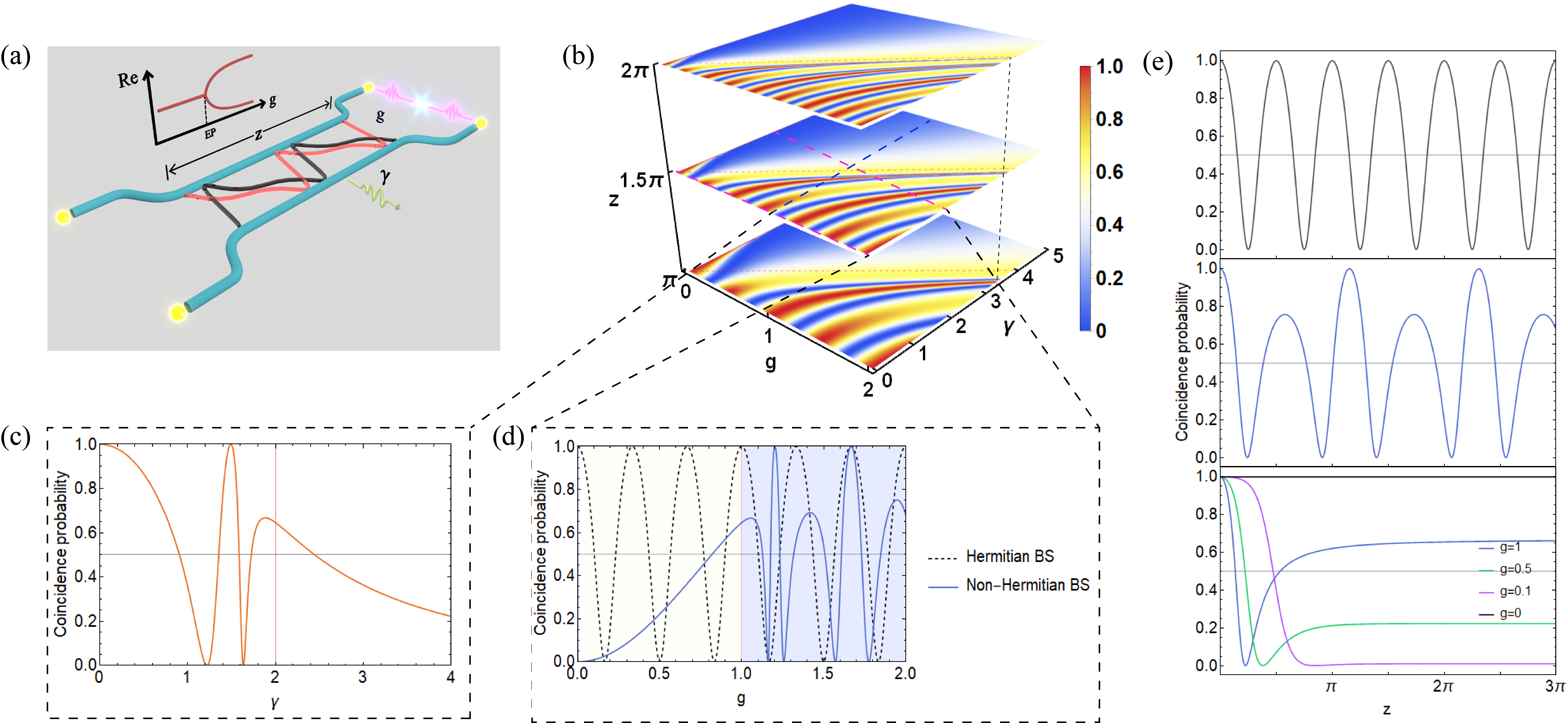}
	\caption{Waveguide setup and coincidence probability analysis near the EP. (a) Waveguide setup exhibiting quantum interference near the EP. (b) Three-dimensional plot of coincidence probability as a function of coupling strength $g$, loss parameter $\gamma$, and propagation length $z$, with the line $g=\frac{\gamma}{2}$ marking the EP. The coincidence probability transitions to narrow peaks as the system approaches the EP with increasing $\gamma$ or $z$ in the PT-symmetry phase. (c) Coincidence probability versus $\gamma$ for $g=1$ and $z=\frac{3}{2}\pi$. (d) Coincidence probability versus $g$ for a unitary beam splitter ($\gamma=0$) and a non-unitary beam splitter ($\gamma=2$) at $z=\frac{3}{2}\pi$, comparing Hermitian and non-Hermitian behaviors. (e) Coincidence probability versus $z$ for a unitary beam splitter ($g=1$, $\gamma=0$, top), a non-unitary beam splitter in the PT-symmetric phase ($g=1$, $\gamma=1$, middle), and the PT-symmetry-broken phase ($\gamma=2$, bottom) with $g=1$, $0.5$, $0.1$, and $0$.}
	\label{fig:g}
\end{figure*}

\noindent \textit{Mathematical model}—Here, we present a general mathematical model for two-photon HOM interference in a non-Hermitian system near an EP. For asymmetric losses in different channels, as shown in Fig. \ref{fig:g}, the Markovian dynamics of the reduced density matrix $\rho(z)$ are described by the standard Lindblad master equation as \cite{lindblad1976generators,klauck2019observation}
\begin{eqnarray}
	\partial_{z}\rho(z)=-\frac{i}{\hbar}[H,\rho]+\gamma(\{a_{2}^{\dagger}a_{2},\rho\}+2a_{2}\rho a_{2}^{\dagger})=\mathcal{L}\rho\label{2},
\end{eqnarray}
where $a_{i}$, $a_{i}^{\dagger}$ represent the annihilation and creation operators, respectively, for the output channel $i$. The unitary evolution between two channels is governed by the Hamiltonian $H=g(a_1^\dagger a_2 + a_1 a_2^\dagger)$ where $g$ is the coupling coefficient. $\gamma$ represents asymmetric loss applied to the second channel, corresponding to the operator $a_{2}$. By postselecting cases where no photons are lost \cite{klauck2025crossing,grafe2013correlations}, the evolution equations for quantum states input into the two channels can be simplified as (see Appendix A and supplementary materials S1 \cite{SupplementalMaterial})
\begin{eqnarray}\label{4}
	U\begin{pmatrix} 1 \\ 0	\end{pmatrix}&=&\csc\frac{\theta}{2}\begin{pmatrix} \sin \frac{\theta+gz\theta_{\lambda}}{2} \\ -i \sin \frac{gz\theta_{\lambda}}{2} \end{pmatrix}\nonumber\\
	U\begin{pmatrix} 0 \\ 1	\end{pmatrix}&=&\csc\frac{\theta}{2}\begin{pmatrix}	-i \sin \frac{gz\theta_{\lambda}}{2} \\ \sin \frac{\theta-gz\theta_{\lambda}}{2} \end{pmatrix},
\end{eqnarray}
where $U$ is the non-unitary evolution operator and $z$ is the propagation distance. We define the dimensionless eigenvalue and eigenvector coalesce coefficient  $\theta_{\lambda}=\sqrt{4- \frac{\gamma^{2}}{g^{2}}} $ and  $\sin\frac{\theta}{2}=\frac{\theta_{\lambda}}{2}$. With the introduction of loss, both $\theta_{\lambda}$ and $\theta$ decrease from 2 and $\pi$ (respectively, as in the Hermitian system) to 0 at the EP, where the eigenvalues and eigenvectors coalesce. Beyond the EP, both $\theta_{\lambda}$ and $\theta$ become imaginary, signaling the transition to the PT-symmetry-broken phase. Introducing loss does not alter the relative $-i$ phase relationship between the transmission and reflection coefficients of $U$, but it causes a change in amplitude offset and a shift in the period. These changes will determine the interference between the two input channels.

Considering HOM interference, where two photons are input from different channels, the normalized coincidence probability for detecting one photon in each output mode, using Eq. \ref{4}, can be written as (see Appendix B and supplementary materials S2 \cite{SupplementalMaterial})
\begin{eqnarray}
	P_{\text{coin}}&=&\frac{\left| u_{11} u_{22} + u_{21} u_{12} \right|^2}{
		\left| u_{11} u_{22} + u_{21} u_{12} \right|^2
		+ \left| u_{11} u_{21} \right|^2
		+ \left| u_{12} u_{22} \right|^2}\nonumber\\
    &=&\frac{1}{1+x} \nonumber\\
    x&=&\frac{4 \left( 1 - \cos \theta \cos(gz \theta_{\lambda}) \right) \sin^2 \left( \frac{gz \theta_{\lambda}}{2} \right)}{\left( 1 + \cos \theta - 2 \cos(gz \theta_{\lambda}) \right)^2}\label{9},
\end{eqnarray}
where $ u_{ij} $s are the matrix elements of $U$.
\\
\hfill

\begin{figure*}[htbp]
	\centering
	\includegraphics[width=7in]{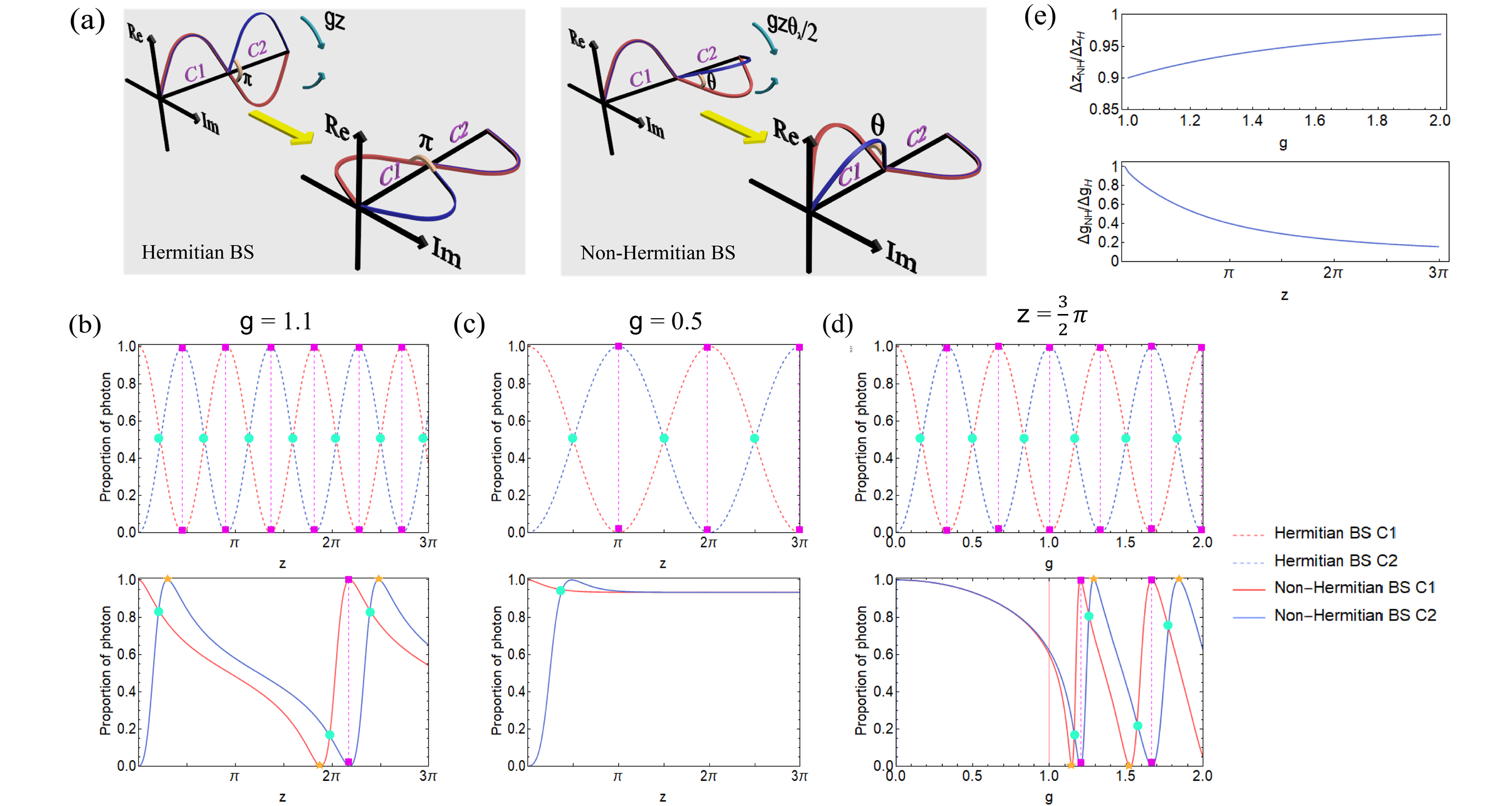}
	\caption{(a) Evolution of the two eigenvectors in the complex plane, showing amplitude distribution between the normal channel (C1) and lossy channel (C2). Red and blue curves trace the two eigenvectors. The amplitude separation between them reflects the relative photon population in each channel. The blue eigenvector rotates clockwise, while the red rotates counterclockwise due to opposite real parts of the eigenvalues. The transition from the lossy channel (before the yellow arrow) to the normal channel (after the arrow) demonstrates the competition between slowed eigenvalue-induced phase accumulation and the shrinking phase-alignment window near the exceptional point, a hallmark of non-Hermitian dynamics. (b, c, d) Photon proportion as a function of propagation length $z$ and coupling strength $g$. The first row shows results for a conventional unitary beam splitter, and the second row shows a non-unitary beam splitter ($\gamma=2$). Green circles, photon bunching events;	Purple squares, anti-bunching events; Orange stars, anti-bunching events that vanish in the non-Hermitian case because loss prevents perfect reflection of both photons. (b) PT-symmetric phase ($g=1.1$) with varying $z$. (c) PT-symmetry-broken phase ($g=0.5$) with varying $z$. (d) Varying $g$ at $z=\frac{3}{2}\pi$. (e) Numerical results for the ratios of spacing in $z$ and $g$ between an anti-bunching and the adjacent bunching point, $\Delta z_{\text{NH}}$ and $\Delta g_{\text{NH}}$, relative to those in the Hermitian system $\Delta z_{\text{H}}$ and $\Delta g_{\text{H}}$, respectively.  }
	\label{fig:gz}
\end{figure*}
\noindent \textit{Result and physical progress—}Fig. \ref{fig:g} illustrates the coincidence transitions between bunching ($P_{\text{coin}}=0$) and anti-bunching ($P_{\text{coin}}=1$). Unlike the cosine oscillations in a Hermitian beam splitter, the non-Hermitian system exhibits non-periodic oscillations in the PT-symmetric phase in both $g$ and $z$, as shown in Fig. \ref{fig:g}(d, e). The number of anti-bunching points is halved compared to bunching points, with some intermediate anti-bunching points vanishing, forming groups where each anti-bunching point is flanked by two bunching points. The interval in $g$ between the two bunching points within each group is much shorter than that across neighboring groups, leading to counterintuitive sharp fluctuations between bunching and anti-bunching within each group. These sharp fluctuations become more pronounced with increasing $z$ and $\gamma$, and intensify near the EP, as evident in Fig. \ref{fig:g}. In the PT-symmetry-broken phase, oscillations between bunching and anti-bunching disappear; coincidence probability varies slowly with $g$ and stabilizes for large $z$, as depicted in Fig. \ref{fig:g}(e).

To achieve a precise mathematical description of the interference behavior, we first examine the coincidence oscillations in the PT-symmetric phase near the EP for two-photon interference. Anti-bunching occurs when two photons, initially injected into different input ports, are detected in separate output ports—this corresponds to both photons being either fully transmitted or fully reflected during their evolution. Combining this physical picture with the analytic expressions in Eqs.~\ref{4} and \ref{9}, the anti-bunching condition can be compactly written as  
\begin{eqnarray}
	\sin \frac{gz\theta_{\lambda }}{2}=0 \label{12} \; \; \text{or} \; \; \sin \frac{\theta+gz\theta_{\lambda}}{2}=\sin \frac{\theta-gz\theta_{\lambda}}{2}=0\label{13} .
\end{eqnarray}
The first condition is satisfied when $gz\theta_{\lambda}=2n\pi$ yielding perfect transmission for both photons in Hermitian and PT-symmetric beam splitters alike. The second condition requires $\theta=\pi$ (i.e., a conventional lossless beam splitter) and $2gz=2n\pi+\pi$ ($n\in\mathbb{N}$), which produces coincidence peaks via perfect reflection of both photons. While full transmission of both photons is possible in PT-symmetric systems, loss completely blocks perfect reflection of both photons. Therefore, anti-bunching caused by reflection disappears, leaving fewer anti-bunching points than in a lossless Hermitian beam splitter.

Bunching points occur when the total amplitude for all anti-bunching events cancels out, as described by the condition $\left| u_{11} u_{22} + u_{21} u_{12} \right|=0$ in Eq. \ref{9}. Due to a relative phase of $-i$ between transmitted and reflected photons, this condition simplifies to equal product of probabilities for two transmissions and two reflections after the evolution process. Therefore, we can simplify the condition to (see Appendix B)
\begin{eqnarray}
	\cos(gz \theta_{\lambda})=\cos^{2} \frac{\theta}{2}\label{15}.
\end{eqnarray}
Apparently, solutions for bunching points are symmetric around $gz\theta_{\lambda}=2n\pi$. For a conventional beam splitter with $\theta=\pi$, these points occur at $2gz=2n\pi\pm\frac{\pi}{2}$, centered between anti-bunching points. As introduction of loss drives the system toward the EP, $\theta$ approaches zero, shifting solutions for $gz\theta_{\lambda}$ closer to $2n\pi$. This results in the formation of groups where each anti-bunching point is flanked by two bunching points, exhibiting rapid fluctuations. Unlike anti-bunching points, which are reduced due to forbidden conditions, bunching points remain unaffected and double in number compared to anti-bunching points.

To further explore the underlying physics and unique quantum optical phenomena associated with these coincidence oscillations, we analyze the photons evolution between two waveguides, as shown in Fig. \ref{fig:gz}. Introducing asymmetric loss drives the system toward the EP, where both eigenvalue splitting $\theta_{\lambda}$ and eigenvector phase separation $\theta$ approach zero. Typically, reduced absolute eigenvalue slows the phase accumulation rate from $gz$ to $\frac{gz\theta_{\lambda}}{2}$, thereby prolonging the oscillation period (see Eq.~\ref{11} in Appendix A). However, near the EP, the eigenvector phases $\pm e^{i\theta}$ nearly coalesce, dramatically compressing the phase accumulation required for photon transfer from the lossy to the normal channel, from $\pi$ to $\theta $. This occurs within narrow intervals $gz\theta_{\lambda}\in[2n\pi-\theta, 2n\pi+\theta]$, shortening the effective transfer distance from $\frac{\pi}{gz}$ to $\frac{2\theta}{gz\theta_{\lambda}}$ and thus producing a sharper trajectory slope, as shown in Fig.~\ref{fig:gz}(a). This effect reshapes the dynamics such that the global oscillation period lengthens, yet rapid channel switching concentrates in narrow windows, yielding strongly asymmetric transfer. This asymmetry breaks reflection-point symmetry between waveguides, suppressing half the anti-bunching peaks in two-photon interference (purple vs. orange markers, Fig.~\ref{fig:gz}(b, d)).

A more significant consequence of near-coalescence eigenvector phase is the reduction of the offset in $gz$ between two transition trajectories, with the scale set by $\theta$. This dual mechanism, steeper trajectory slopes combined with the reduced offset, profoundly alters the quantum interference pattern: bunching events become confined to the narrow windows (green markers, Fig.~\ref{fig:gz}(b, d)), effectively sandwiching each anti-bunching point ($gz\theta_{\lambda} = 2n\pi$) between two closely spaced bunching points. The resulting tight groups exhibit extremely rapid intra-group transitions that become progressively steeper as the EP is approached, as shown in Fig.~\ref{fig:gz}(d). While the coincidence variation in $z$ is slightly sharper than in Hermitian systems, consistent with prior observations for the first bunching point \cite{klauck2019observation,zhou2022characterization}, the sensitivity to coupling strength $g$ can be much more pronounced, as shown in Fig. \ref{fig:g}(d, e). This heightened sensitivity arises because both $\theta_{\lambda}$ and $\theta$ depend on $g$. The spacing in $z$ between an anti-bunching and the adjacent bunching point is $\Delta z_{\text{NH}}=\frac{\arccos(1-\theta_{\lambda}^{2}/4)}{g\theta_{\lambda}}$. Compared to the corresponding Hermitian spacing $\Delta z_{\text{H}}$, the ratio $\Delta z_{\text{NH}}/\Delta z_{\text{H}}$ remains bounded between 1 and approximately $\frac{2\sqrt{2}}{\pi}$. In contrast, the spacing in $g$, governed by the transcendental relation in Eq.~\ref{15}, allows the ratio $\Delta g_{\text{NH}}/\Delta g_{\text{H}}$ to range from 1 down to nearly 0 with the increase in $z$. This much wider dynamic range explains the dramatically increased sensitivity to $g$, as shown in Fig. \ref{fig:gz}(e).

\begin{figure*}[htbp]
	\centering
	\includegraphics[width=7in]{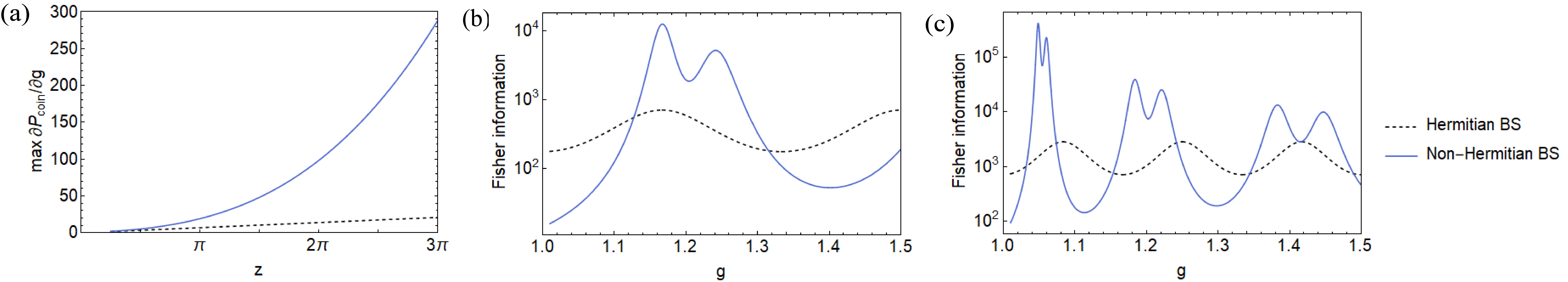}
	\caption{Sensing capability of a conventional unitary beam splitter (gray dashed line) versus a non-unitary beam splitter with single-channel loss (blue solid line). (a) Maximum slope of coincidence probability $P_{\text{coin}}$ with respect to coupling strength $g$ as a function of propagation length $z$ in the PT-symmetric phase for $\gamma=2$. (b, c) Fisher information at $z=\frac{3}{2}\pi$ (b) and $z=3\pi$ (c), quantifying sensitivity.}
	\label{fig:g1}
\end{figure*}

We further demonstrate the enhanced sensing capability for $g$ in Fig. \ref{fig:g1}. As shown in Fig. \ref{fig:g1}(a), the slope of the coincidence probability with respect to coupling strength $g$ in the non-Hermitian beam splitter significantly exceeds that of the Hermitian case. To further quantify sensitivity, we use the Fisher information per recorded trial (see supplementary materials S4 \cite{SupplementalMaterial})
\begin{eqnarray} 
	F = \left\langle \left( \frac{\partial \ln(p_i)}{\partial g} \right)^2 \right\rangle ,
\end{eqnarray}
where $i \in \{02, 20, 11\}$, which is inversely proportional to the mean square error \cite{giovannetti2011advances,slussarenko2017unconditional}. As shown in Fig. \ref{fig:g1}(b, c), the Fisher information near the EP exhibits orders-of-magnitude advantage over Hermitian systems. This heightened sensitivity, far surpassing that of a conventional beam splitter, enables the non-Hermitian system to serve as a highly effective quantum sensor. By leveraging the sharp transitions in quantum interference, particularly near the exceptional point, this system can detect minute changes in coupling strength $g$ with exceptional precision. Such enhanced sensitivity makes it promising for applications in quantum sensing, including high-resolution measurements of environmental perturbations and material properties, offering significant advantages over traditional Hermitian systems.
\\
\hfill

Next, we consider the coincidence variation in the PT-symmetry-broken phase. In this phase, the parameters $\theta_{\lambda} $ and $\theta $ become purely imaginary, with $\theta_{\lambda}=i\theta^{'}_{\lambda}$, where $\theta^{'}_{\lambda}=2\sqrt{(\frac{\gamma}{2g})^2-1}$. This eliminates the periodic behavior of transmission and reflection, disrupting the periodicity of bunching and anti-bunching. This phenomenon can be analyzed using the bunching and anti-bunching conditions given in Eqs. \ref{12} and \ref{15}. Specifically, the only possible anti-bunching condition that can be satisfied is $\sinh \frac{gz\theta^{'}_{\lambda }}{2}=0 $, which occurs only when $g=0$ or $z=0$. This is expected, as the absence of coupling or evolution should always lead to anti-bunching. In contrast, the bunching condition $  \cosh (gz\theta^{'}_{\lambda})=\cos^{2} \frac{\theta}{2}$ always yields a unique solution for photon transmission $z=\frac{1}{g \theta_{\lambda}^{'}}\operatorname{arccosh}(1-\frac{\theta_{\lambda}^{'2}}{4})<\frac{\pi}{4g}$ due to the monotonic nature of the $\cosh x$. As $z$ increases further, typically beyond $\pi$, the transmission and reflection probabilities converge, causing the coincidence rate to approach a stable value which can be tuned by adjusting $g$, as shown in Fig. \ref{fig:g}(d, e).

In the PT-symmetry-broken phase, as coupling drops significantly below loss, photons predominantly remain in the normal channel over the lossy channel. Regardless of the initial channel, the photon distribution converges to a stable value near the normal channel, yielding a $z$-independent stable coincidence probability, as shown in Fig. \ref{fig:gz}(c). Interestingly, the dual mechanism cause by $\theta$ also exists in the first period like the PT-symmetric situation, and produces a crosspoint corresponding to a bunching point within $gz|\theta_{\lambda}|\in[0, |\theta|]$.

To further gain the mathematical description for the stable coincidence probability, we consider the regime where $z$ is large enough that $\cosh(gz\theta^{'}_{\lambda})\gg1$. In this limit, $x$ can be approximated as (see supplementary materials S5 \cite{SupplementalMaterial})
\begin{eqnarray}
	x\approx\dfrac{2+ \theta_{\lambda }^{'2} }{4} \label{16}.
\end{eqnarray}
Therefore, $P_{\text{coin}}$ will approach a constant value, $ \frac{4}{6+ \theta_{\lambda }^{'2}}$, for a given $\theta_{\lambda }^{'} $. As shown in Fig. \ref{fig:g}(e), $z\geq\pi$ is sufficient to achieve this stable value. This unique property enables a non-Hermitian beam splitter to maintain a stable coincidence probability between 0 and $\frac{2}{3}$, which persists over propagation length $z$ and can be tuned by adjusting the coupling parameter $g$, in contrast to the oscillatory behavior of Hermitian systems. This stability offers significant advantages for quantum technologies, enabling reliable and controllable two-photon interference for applications such as quantum information processing, where consistent coincidence rates are crucial for quantum state preparation and manipulation. Additionally, the ability to tune $P_{\text{coin}}$ via $g$ allows for precise control in quantum communication protocols, enhancing the robustness of quantum key distribution and entanglement distribution. Note that when $g\rightarrow0$, the simplified model does not apply, and a sharp transition in the coincidence probability emerges (see Appendix D).
 \\
\hfill

Finally, we examine the system crossing the EP. As parameters approach the EP, the Hamiltonian becomes defective, causing coalescence of eigenvalues and eigenvectors, make challenging to analyze. However, for $|\theta|\ll gz|\theta_{\lambda}|$, we can approximate the evolution operator as
\begin{eqnarray}
	U_{\text{EP}}= \sigma_{z}-i\sigma_{x},
\end{eqnarray}
where $\sigma_{i} $ is the Pauli matrix. This evolution operator yields coincidence results independent of parameter choices, giving $x=\frac{1}{2}$ and a coincidence probability of $P_{\text{coin}}=\frac{2}{3}$. This confirms the EP limit in Eqs. \ref{16} and \ref{14} (in End matter). Consequently, the transition crossing point with respect to $g$ converges to a specific value, and the stable coincidence probability remains largely invariant with $z$, as shown in Fig. \ref{fig:g}(d, e). Despite the solution's singularity, the EP crossing is typically smooth \cite{longhi2018quantum,klauck2025crossing}.

In this work, we have demonstrated an EP-induced transition in quantum interference within a non-Hermitian coupled waveguide system. The coincidence oscillation pattern transitions from highly sensitive to robust behavior upon crossing the EP, in sharp contrast to the uniformly stable behavior of Hermitian systems. Through theoretical analysis, we reveal that the local dual mechanism caused by the near-coalescence phase of eigenvectors in the PT-symmetric phase drives rapid, counterintuitive transitions between photon bunching and anti-bunching. Conversely, in the PT-symmetry-broken phase, imaginary eigenvalues result in a coincidence rate that is remarkably insensitive to parameter changes. These properties enable compelling applications. The steep interference slope in the PT-symmetric phase enhances sensitivity for sensing, surpassing Hermitian systems. Meanwhile, the PT-symmetry-broken phase offers stable, tunable two-photon coincidence outcomes, ranging from partial anti-bunching to near bunching, controlled by coupling parameters. These capabilities suggest significant potential for non-Hermitian systems in quantum information technologies, including quantum communication, computation, and high-precision sensing.

	\begin{acknowledgments}
		We acknowledge useful conversations with Sixin Chen and Wange Song. The work at The University of Hong Kong was sponsored by the New Cornerstone Science Foundation, the Research Grants Council of Hong Kong (AoE/P-502/20, STG3/E-704/23-N, 17309021), the Guangdong Provincial Quantum Science Strategic Initiative (GDZX2204004, GDZX2304001). 		
	\end{acknowledgments}

	\nocite{*}
	\bibliographystyle{ieeetr}
\bibliography{article}

\clearpage

\appendix
\onecolumngrid
\section*{\label{sec:leveled}End Matter}	

\twocolumngrid
\subsection*{\label{sec:leveled1}Appendix A--Evolution operator for a non-Hermitian beam splitter }

We use postselection to focus on cases where both photons remain, as shown in Eq.~\ref{2}. This is often studied in experiments to explore quantum interference, like bunching or anti-bunching. Postselection removes quantum jump terms, leading to dynamics driven by the effective non-Hermitian Hamiltonian $H_{\text{eff}}$ as

\begin{eqnarray}
	H_{\text{eff}}	 &=&H-\gamma a_{2}^{\dagger}a_{2}=\begin{pmatrix} i \frac{\gamma}{2} & g \\
		g & -i \frac{\gamma}{2} \end{pmatrix}	-  i \frac{\gamma}{2} \begin{pmatrix} 1 &0 \\
		0 & 1 \end{pmatrix}\nonumber\\&=&H_{0}	-  i \frac{\gamma}{2} I_{2},
\end{eqnarray}
where the common loss term $i\frac{\gamma}{2}I_2$ can be factored out when analyzing photon interference, allowing us to focus on the EP-dominant term $H_0$. The evolution of the states can be viewed as a superposition of the evolutions of the eigenvectors, such that

\begin{eqnarray}
	U\begin{pmatrix} 1 \\ 0	\end{pmatrix}&=&\frac{i e^{\frac{i \theta}{2}}}{ 1-e^{i \theta}} \left( e^{-i \lambda_{1}z}\ket{\varphi_{1}} -e^{-i \lambda_{2}z}\ket{\varphi_{2}}\right)\nonumber\\
	&=&\csc\frac{\theta}{2}\begin{pmatrix} \sin \frac{\theta+gz\theta_{\lambda}}{2} \\ -i \sin \frac{gz\theta_{\lambda}}{2} \end{pmatrix}\nonumber\\
	U\begin{pmatrix} 0 \\ 1	\end{pmatrix}&=&\frac{1}{ 1-e^{i \theta}} \left( e^{-i \lambda_{1}z} \ket{\varphi_{1}} - e^{-i \lambda_{2}z} e^{i \theta} \ket{\varphi_{2}}\right)\nonumber\\
	&=&\csc\frac{\theta}{2}\begin{pmatrix}	-i \sin \frac{gz\theta_{\lambda}}{2} \\ \sin \frac{\theta-gz\theta_{\lambda}}{2} \end{pmatrix}\label{11},
\end{eqnarray}
where $ \lambda_{i}$ and $\ket{\varphi_{i}} $ are the eigenvalues and eigenvectors of $H_0$, respectively. $z$ is the propagation distance. Thus, we obtain that the evolution operator satisfies

\begin{eqnarray}
	U= \csc\frac{\theta}{2}\begin{pmatrix} \sin \frac{\theta+gz\theta_{\lambda}}{2} & -i \sin \frac{gz\theta_{\lambda}}{2} \\	-i \sin \frac{gz\theta_{\lambda}}{2} & \sin \frac{\theta-gz\theta_{\lambda}}{2} \end{pmatrix} .
\end{eqnarray}

We can see that introducing loss preserves the relative $-i$ phase between transmission and reflection coefficients but alters their amplitude offset and period. In a Hermitian system, these coefficients follow cosine and sine functions with identical phases. At the EP, where $\theta_{\lambda}$ and $\theta$ approach zero, the solution becomes singular, and the amplitudes of the coefficients fully overlap. For sufficiently large $z$, such that $|\theta|\ll gz|\theta_{\lambda}|$, the term $\sin \frac{\theta - gz\theta_{\lambda}}{2}$ approximates to $-\sin \frac{gz\theta_{\lambda}}{2}$, introducing an additional $-1$ phase factor. Consequently, the evolution operator is modified as

\begin{eqnarray}
	U_{\text{EP}}= \begin{pmatrix} 1 & -i \\	-i & -1 \end{pmatrix}= \sigma_{z}-i\sigma_{x}.
\end{eqnarray}

\subsection*{\label{sec:leveled2}Appendix B--Bunching and anti-bunching conditions}

Considering HOM interference, where two photons are input from different channels, the coincidence intensity can be represented as

\begin{eqnarray}
	I_{\text{coin}}=\left| u_{11} u_{22} + u_{21} u_{12} \right|^2,
\end{eqnarray}
where $ u_{ij} $s are the matrix elements of $U$. Considering the normalized coefficient, the normalized coincidence probability $P_{\text{coin}} $ can be written as

\begin{eqnarray}
	P_{\text{coin}}&=&\frac{I_{\text{coin}}}{I_{\text{total}}}=\frac{\left| u_{11} u_{22} + u_{21} u_{12} \right|^2}{
		\left| u_{11} u_{22} + u_{21} u_{12} \right|^2
		+ \left| u_{11} u_{21} \right|^2
		+ \left| u_{12} u_{22} \right|^2}\nonumber\\
	&=&\frac{1}{1+x}.
\end{eqnarray}
By comparing the solution of the evolution operator $U$, we can solve for $x$ as presented in Eq \ref{9}. Here, we can see that the normalized coincidence probability $P_{\text{coin}}$ ranges from 0 to 1. When anti-bunching occurs, $P_{\text{coin}}=1$, which corresponds to $x=0$. Conversely, when bunching occurs, $P_{\text{coin}}=0$, and this corresponds to $x\rightarrow\infty$. The value $P_{\text{coin}}=0.5$ represents the classical case where there is no quantum interference, indicating that the photons are completely distinguishable.

The anti-bunching condition, given by $x=0$, can be expressed as:

\begin{eqnarray}
	\left( 1 - \cos \theta \cos(gz \theta_{\lambda}) \right) \sin^2 \left( \frac{gz \theta_{\lambda}}{2}\right)=0.
\end{eqnarray}
This condition can be divided into two subconditions: $\sin^2 \left( \frac{gz \theta_{\lambda}}{2} \right) = 0$ and $1 - \cos \theta \cos(gz \theta_{\lambda}) = 0$. Using the non-unitary operator $U$, these can be further simplified in terms of the transmission and reflection coefficients as
\begin{eqnarray}
	\sin \frac{gz\theta_{\lambda }}{2}=0 \; \text{or} \; \sin \frac{\theta+gz\theta_{\lambda}}{2}=\sin \frac{\theta-gz\theta_{\lambda}}{2}=0\label{} .
\end{eqnarray}

The bunching condition, characterized by equal product of probabilities for two transmissions and two reflections after the evolution process, corresponds to the limit $x\rightarrow\infty$ and can be expressed as

\begin{eqnarray}
	\sin^{2} \frac{gz\theta_{\lambda }}{2}=\sin \frac{\theta+gz\theta_{\lambda}}{2}\sin \frac{\theta-gz\theta_{\lambda}}{2}.
\end{eqnarray}
This condition can then be simplified as
\begin{eqnarray}
	\cos(gz \theta_{\lambda})=\cos^{2} \frac{\theta}{2}.
\end{eqnarray}

\subsection*{\label{sec:leveled3}Appendix C--Taylor expansion for the lineshape in the PT-symmetry phase }

\begin{figure}[htbp]
	\centering
	\includegraphics[width=3in]{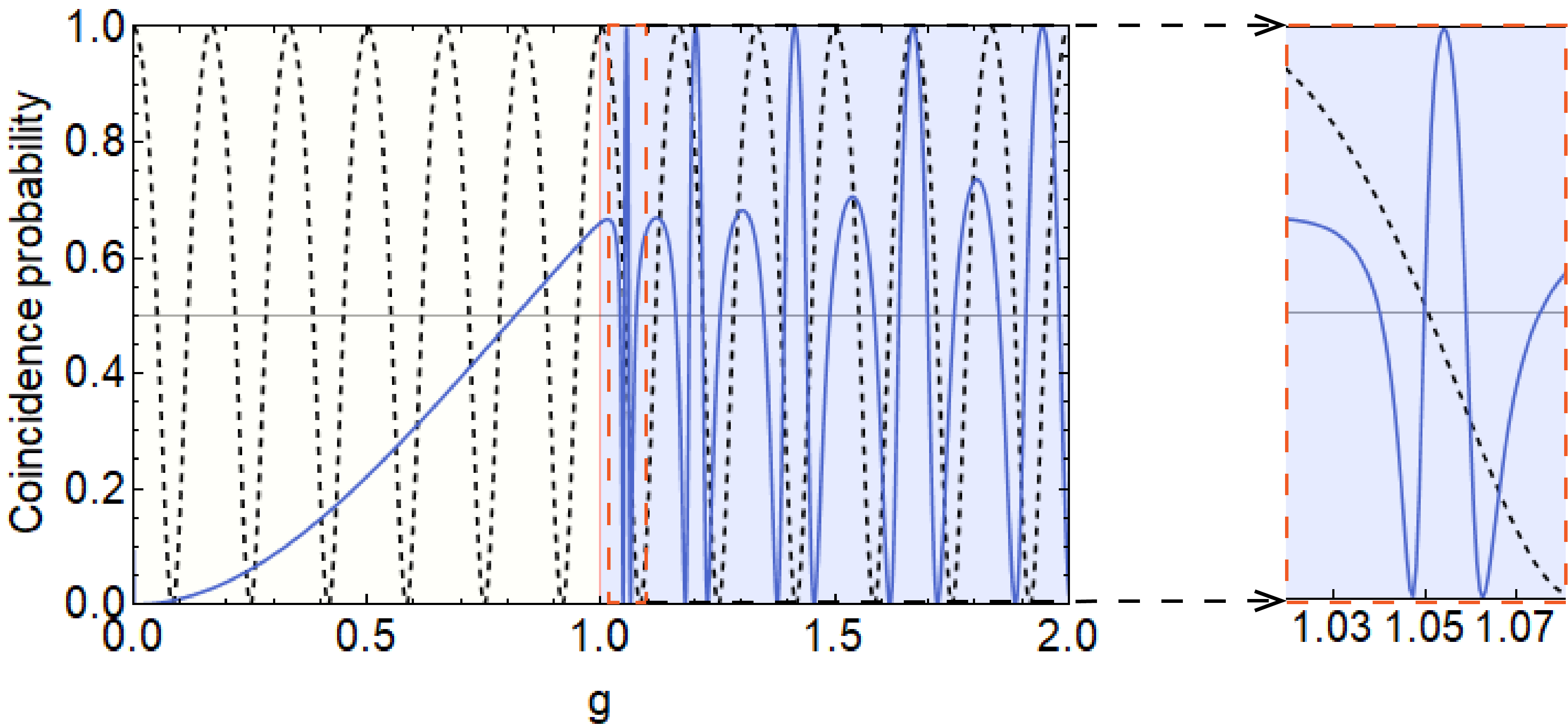}
	\caption{ Coincidence probability variation with $g$ at $z=3\pi$ for a conventional unitary beam splitter (gray dashed line) versus a non-unitary beam splitter with single-channel loss (blue solid line). }
	\label{fig:gz3}
\end{figure}

To analyze $P_{\text{coin}}$ in EP limit and the intra-group lineshape in the PT-symmetry phase, we can also perform a Taylor expansion around the anti-bunching point $\theta_{\lambda 0}=\dfrac{2n\pi}{gz} $, with $\theta_{\lambda}=\theta_{\lambda 0}+\delta $ where $\delta\ll \theta_{\lambda 0}$. In this case, $x$ can be approximated as (see supplementary materials S3 \cite{SupplementalMaterial})
\begin{eqnarray}
	x\approx\frac{\left( 1 - k \right) \varepsilon^4 +2k\varepsilon^{2}}{2 \left( \varepsilon^{2}-k \right)^2}\label{14},
\end{eqnarray}
where we define $k= \frac{\theta_{\lambda 0}^{2}}{2}$ and $\varepsilon=g z \delta$. At the zero point $x=0$, where $\delta=0$, the coincidence probability $P_{\text{coin}}=1$ indicates an anti-bunching point. As $|\delta|$ increases, $x$ grows rapidly, diverging as $|\delta|\rightarrow\frac{\sqrt{k}}{gz}$, where $P_{\text{coin}}=0$, corresponding to a bunching point. For $|\delta|\gg\frac{\sqrt{k}}{gz}$, $x$ approaches a constant value $\frac{1-k}{2}$, and $P_{\text{coin}}=\frac{2}{3-k}$ reaches $\frac{2}{3}$ near the EP when $gz\theta_{\lambda}\gg\theta$. The interval $\frac{\sqrt{k}}{gz}$ determines the fluctuation range and thus the intra-group transition slope. Approaching the EP by reducing $k$ increases the transition slope, and increasing $z$ also produces sharper, more distinct peaks, as depicted in Fig. \ref{fig:gz3} for $z=3\pi$.

\subsection*{\label{sec:leveled4}Appendix D--Coincidence probability switch when $g\rightarrow0$ }

\begin{figure}[htbp]
	\centering
	\includegraphics[width=3.7in]{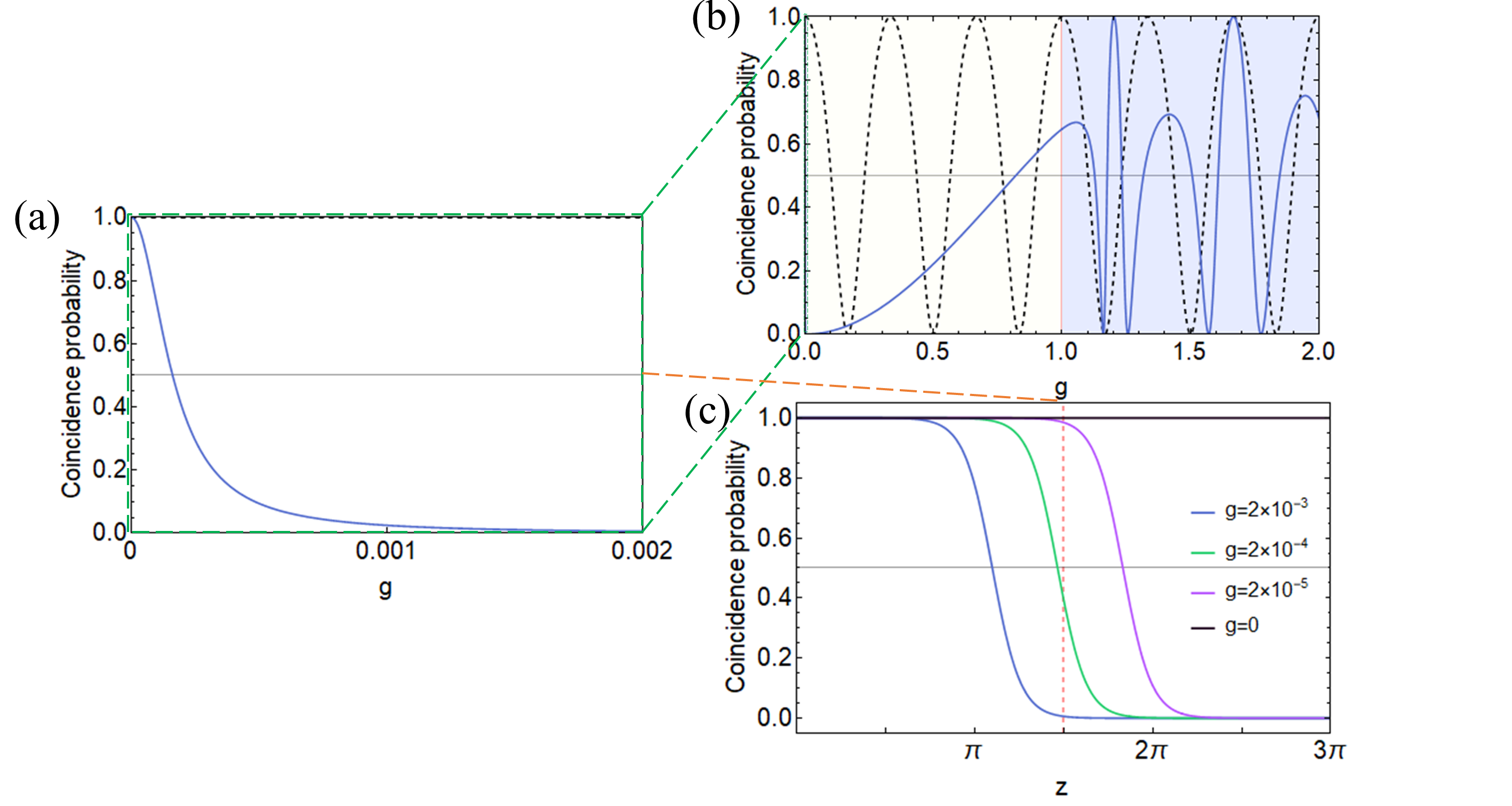}
	\caption{ Coincidence probability variation when $g\rightarrow0$. (a, b) show the coincidence probability variation with $g$ at $z=\frac{3}{2}\pi$ for a conventional unitary beam splitter (gray dashed line) versus a non-unitary beam splitter with single-channel loss (blue solid line). (c) shows the coincidence probability variation with $z$ for different $g$ when $g\rightarrow0$. }
	\label{fig:g0}
\end{figure}

In the PT-symmetry-broken phase, the coincidence probability generally exhibits stability, but a sharp transition from bunching to anti-bunching emerges as $g\rightarrow0$, as shown in Fig.~\ref{fig:g}(d, e) and \ref{fig:g0}. This arises because the first-period transition persists in both PT-symmetric and PT-broken phases, as illustrated in Fig.~\ref{fig:g1}(b, c). When $gz$ is sufficiently large, the coincidence probability stabilizes in the PT-broken phase. However, as $g$ approaches zero, photon transfer between channels becomes increasingly difficult, requiring greater propagation distance $z$. This causes the first-period transition to dominate. As shown in Fig.~\ref{fig:g0}, gradually reducing $g$ shifts this transition to larger $z$. Due to the high sensitivity of this shift to $g$, a rapid transition from bunching to anti-bunching is observed.

\clearpage

\onecolumngrid
\setcounter{equation}{0}
\setcounter{figure}{0}
\setcounter{subsection}{0}
\renewcommand{\theequation}{S.\arabic{equation}}
\renewcommand{\thefigure}{S\arabic{figure}}
\renewcommand{\thesubsection}{S\arabic{subsection}}

\section*{Supplementary materials}

\subsection{\label{sec:level51}Derivation of evolution operator for a non-Hermitian beam splitter }

Here we consider a passive non-Hermitian system with two channels, described by the input operator $\begin{pmatrix} a_1 \\ a_2 \end{pmatrix}$. The coupling between channels is governed by the Hamiltonian $H=g(a_1^\dagger a_2 + a_1 a_2^\dagger)$, with a Markovian loss $\gamma$ in the second channel. The evolution of the density matrix $\rho(z)$ in the waveguide is described by the standard Lindblad master equation as

\begin{eqnarray}
	\partial_{z}\rho(z)&=&-\frac{i}{\hbar}[H,\rho]+\gamma(\{a_{2}^{\dagger}a_{2},\rho\}+2a_{2}\rho a_{2}^{\dagger})\nonumber\\&=&-i(H_{\text{eff}}\rho-\rho H_{\text{eff}}^{\dagger})+2\gamma a_{2}\rho a_{2}^{\dagger}\label{1}.
\end{eqnarray}
We apply postselection to consider only cases where both photons remain, a situation typically examined in experiments to study quantum interference for bunching or anti-bunching. This postselection projects out quantum jump terms, resulting in dynamics governed by the effective non-Hermitian Hamiltonian $H_{\text{eff}}$ as

\begin{eqnarray}
	H_{\text{eff}}	 &=&\begin{pmatrix} 0 & g \\
		g & -i \gamma	\end{pmatrix} 
	=\begin{pmatrix} i \frac{\gamma}{2} & g \\
		g & -i \frac{\gamma}{2} \end{pmatrix}	-  i \frac{\gamma}{2} \begin{pmatrix} 1 &0 \\
		0 & 1 \end{pmatrix}\nonumber\\&=&H_{0}	-  i \frac{\gamma}{2} I_{2},
\end{eqnarray}
where the common loss term $i\frac{\gamma}{2}I_2$ can be factored out when analyzing photon interference, allowing us to focus on the EP-dominant term $H_0$. The eigenvalues and eigenvectors of $H_0$ can then be calculated as

\begin{eqnarray}
	\lambda_{1}&=&-\frac{g\theta_{\lambda}}{2} \qquad \ket{\varphi_{1}}=\begin{pmatrix}
		i e^{i \frac{\theta}{2}} \\ 1	\end{pmatrix}\nonumber\\
	\lambda_{2}&=&\frac{g\theta_{\lambda}}{2}  \qquad \ket{\varphi_{2}}=\begin{pmatrix}
		i e^{-i \frac{\theta}{2}} \\ 1	\end{pmatrix},
\end{eqnarray}
where the dimensionless eigenvalue and eigenvector coalesce coefficient are $\theta_{\lambda}=\sqrt{4- \frac{\gamma^{2}}{g^{2}}} $ and  $\sin\frac{\theta}{2}=\frac{\theta_{\lambda}}{2}$. With the introduction of loss, both $\theta_{\lambda}$ and $\theta$ decrease from 2 and $\pi$ (respectively, as in the Hermitian system) to 0 at the EP, where the eigenvalues and eigenvectors coalesce. Beyond the EP, both $\theta_{\lambda}$ and $\theta$ become imaginary.

Consider the states in which the photons input into the two channels satisfy

\begin{eqnarray}
	\begin{pmatrix} 1 \\ 0	\end{pmatrix}&=& \frac{i e^{\frac{i \theta}{2}}}{ 1-e^{i \theta}} \left( \ket{\varphi_{1}} -\ket{\varphi_{2}}\right)\nonumber\\
	\begin{pmatrix} 0 \\ 1	\end{pmatrix}&=& \frac{1}{ 1-e^{i \theta}} \left(  \ket{\varphi_{1}} -  e^{i \theta} \ket{\varphi_{2}}\right).
\end{eqnarray}
The evolution of the states can be viewed as a superposition of the evolutions of the eigenvectors, such that

\begin{eqnarray}
	U\begin{pmatrix} 1 \\ 0	\end{pmatrix}&=&\frac{i e^{\frac{i \theta}{2}}}{ 1-e^{i \theta}} \left( e^{-i \lambda_{1}z}\ket{\varphi_{1}} -e^{-i \lambda_{2}z}\ket{\varphi_{2}}\right)\nonumber\\
	&=& \frac{i e^{\frac{i \theta}{2}}}{ 1-e^{i \theta}} \left( e^{i \frac{gz \theta_{\lambda}}{2}}\begin{pmatrix}
		i e^{i \frac{\theta}{2}} \\ 1	\end{pmatrix} - e^{-i \frac{gz \theta_{\lambda}}{2}}\begin{pmatrix} i e^{-i \frac{\theta}{2}} \\ 1	\end{pmatrix}\right)\nonumber\\&=&\csc\frac{\theta}{2}\begin{pmatrix} \sin \frac{\theta+gz\theta_{\lambda}}{2} \\ -i \sin \frac{gz\theta_{\lambda}}{2} \end{pmatrix}\nonumber\\
	U\begin{pmatrix} 0 \\ 1	\end{pmatrix}&=&\frac{1}{ 1-e^{i \theta}} \left( e^{-i \lambda_{1}z} \ket{\varphi_{1}} - e^{-i \lambda_{2}z} e^{i \theta} \ket{\varphi_{2}}\right)\nonumber\\
	&=& \frac{1}{ 1-e^{i \theta}} \left(  e^{i \frac{gz \theta_{\lambda}}{2}}\begin{pmatrix}
		i e^{i \frac{\theta}{2}} \\ 1	\end{pmatrix} -e^{-i \frac{gz \theta_{\lambda}}{2}}  e^{i \theta} \begin{pmatrix} i e^{-i \frac{\theta}{2}} \\ 1	\end{pmatrix}\right)\nonumber\\&=&\csc\frac{\theta}{2}\begin{pmatrix}	-i \sin \frac{gz\theta_{\lambda}}{2} \\ \sin \frac{\theta-gz\theta_{\lambda}}{2} \end{pmatrix},
\end{eqnarray}
where $z$ is the propagation distance. Thus, we obtain that the evolution operator satisfies

\begin{eqnarray}
	U= \csc\frac{\theta}{2}\begin{pmatrix} \sin \frac{\theta+gz\theta_{\lambda}}{2} & -i \sin \frac{gz\theta_{\lambda}}{2} \\	-i \sin \frac{gz\theta_{\lambda}}{2} & \sin \frac{\theta-gz\theta_{\lambda}}{2} \end{pmatrix} \label{66}.
\end{eqnarray}

\subsection{\label{sec:level52}Derivation of normalized coincidence probability in HOM interference}

Considering HOM interference, where two photons are input from different channels, the coincidence intensity for detecting one photon in each output mode can be expressed as \cite{scully1997quantum}:

\begin{eqnarray}
	I_{\text{coin}}=\bra{1_{1}1_{2}}a_{4}^{\dagger}a_{3}^{\dagger}a_{3}a_{4}\ket{1_{1}1_{2}}
\end{eqnarray}
where $\ket{1_{1}1_{2}}$ denotes the input state with one photon in input channel 1 and one photon in input channel 2, and $a_{i} $ are the annihilation operators for the output channel $i$. For the evolution of a two-photon input state through a beam splitter, the output annihilation operators $\begin{pmatrix}	a_{3} \\ a_{4}	\end{pmatrix}$ can be written in terms of the input operators  $\begin{pmatrix}	a_{1} \\ a_{2}	\end{pmatrix}$ as

\begin{eqnarray}
	\begin{pmatrix}	a_{3} \\ a_{4}	\end{pmatrix}=U\begin{pmatrix}	a_{1} \\ a_{2}	\end{pmatrix}
\end{eqnarray}
Thus, the coincidence intensity can be represented as

\begin{eqnarray}
	I_{\text{coin}}=\left| u_{11} u_{22} + u_{21} u_{12} \right|^2
\end{eqnarray}
where we define the beam splitter transformation matrix as $U= \begin{pmatrix} u_{11} & u_{12} \\	u_{21} & u_{22} \end{pmatrix}$. Considering the normalized coefficient, the normalized coincidence probability $P_{\text{coin}} $ can be written as

\begin{eqnarray}
	P_{\text{coin}}&=&\frac{I_{\text{coin}}}{I_{\text{total}}}=\frac{\left| u_{11} u_{22} + u_{21} u_{12} \right|^2}{
		\left| u_{11} u_{22} + u_{21} u_{12} \right|^2
		+ \left| u_{11} u_{21} \right|^2
		+ \left| u_{12} u_{22} \right|^2}\nonumber\\
	&=&\frac{1}{1+x} \nonumber\\
	x&=&\frac{ \left| u_{11} u_{21} \right|^2
		+ \left| u_{12} u_{22} \right|^2}{
		\left| u_{11} u_{22} + u_{21} u_{12} \right|^2}
\end{eqnarray}
Comparing the solution of the evolution operator $U$ in Eq. \ref{66}, we can solve for $x$ by

\begin{eqnarray}
	x&=&\frac{ \left| u_{11} u_{21} \right|^2
		+ \left| u_{12} u_{22} \right|^2}{
		\left| u_{11} u_{22} + u_{21} u_{12} \right|^2}\nonumber\\
	&=&	\frac{ \left| \sin \frac{\theta+gz\theta_{\lambda}}{2}  \sin \frac{gz\theta_{\lambda}}{2}\right|^2
		+ \left|  \sin \frac{gz\theta_{\lambda}}{2} \sin \frac{\theta-gz\theta_{\lambda}}{2} \right|^2}{
		\left|\sin \frac{\theta+gz\theta_{\lambda}}{2} \sin \frac{\theta-gz\theta_{\lambda}}{2}- \sin^{2} \frac{gz\theta_{\lambda}}{2}\right|^2}\nonumber\\
	&=&\frac{4 \left( 1 - \cos \theta \cos(gz \theta_{\lambda}) \right) \sin^2 \left( \frac{gz \theta_{\lambda}}{2} \right)}{\left( 1 + \cos \theta - 2 \cos(gz \theta_{\lambda}) \right)^2}
\end{eqnarray}

\subsection{\label{sec:level54}Derivation for Taylor expansion of $x$ in the PT-symmetry phase }

We consider the first two terms of the Taylor expansion of $x$ in Eq. 19 of the main text. We choose the anti-bunching points $\theta_{\lambda 0}=\dfrac{2n\pi}{gz} $, and set $\theta_{\lambda}=\theta_{\lambda 0}+\delta $, where $\delta\ll \theta_{\lambda 0}$. Expanding to second order in $\delta$, we have $\cos(gz \theta_{\lambda})\approx 1 - \frac{(gz\delta)^2}{2} $ and $ \sin\left( \frac{gz \theta_{\lambda}}{2}\right)\approx \frac{gz\delta}{2}$. Therefore, $x$ can be approximated as

\begin{eqnarray}
	x&\approx&\frac{4\left( 1 - (1-\frac{\theta_{\lambda 0}^{2}}{2})  \left( 1 - \frac{(gz \delta)^2}{2} \right) \right) (\frac{gz \delta}{2})^2}{
		\left( 1 +(1-\frac{\theta_{\lambda 0}^{2}}{2}) - 2 \left( 1 - \frac{(gz \delta)^2}{2} \right) \right)^2
	}\nonumber\\
	&=&\frac{g^2\delta^2 z^2 \left( 2 + (-2 + g^2\delta^2 z^2) (1-\frac{\theta_{\lambda 0}^{2}}{2}) \right)}{2 \left(  g^2\delta^2 z^2-\frac{\theta_{\lambda 0}^{2}}{2} \right)^2}
	\nonumber\\
	&=&\frac{\left( 1 - k \right) \varepsilon^4 +2k\varepsilon^{2}}{2 \left( \varepsilon^{2}-k \right)^2}
\end{eqnarray}
where we define $k= \frac{\theta_{\lambda 0}^{2}}{2}$ and $\varepsilon=gz\delta $.

\subsection{\label{sec:level56}Fisher information }

For a sensing system estimating parameter $g$ with probability function $p_i(g)$ for measurement events $i$, the precision is limited by the Cramér-Rao bound \cite{kay1993fundamentals}, which states that the variance of any unbiased estimator satisfies
\begin{eqnarray} \Delta^2 g \geq \frac{1}{N F} \end{eqnarray}
where $N$ is the repeat times of the recorded trial, and the $F$ is the Fisher Information, expressed as
\begin{eqnarray} F = \sum_i \left( \frac{\partial p_i}{\partial g} \right)^2 \frac{1}{p_i} = \left\langle \left( \frac{\partial \ln(p_i)}{\partial g} \right)^2 \right\rangle \end{eqnarray}
Larger Fisher Information indicates higher estimation precision.

\subsection{\label{sec:level55}Derivation for $x$ in the PT-symmetry-broken phase}

Here, we consider the derivation of $x$ in Eq. 7 of the main text. For sufficiently large $z$ such that $\cosh (gz\theta^{'}_{\lambda})\gg1$, we can approximate $x$ as

\begin{eqnarray}
	x&=&\dfrac{4 \left( 1 - \cos \theta \cosh(gz \theta^{'}_{\lambda}) \right)\sinh^2 \left( \frac{gz \theta^{'}_{\lambda}}{2} \right) }{\left( 1 + \cos \theta - 2 \cosh(gz \theta^{'}_{\lambda}) \right)^2}\nonumber\\
	&\approx&\dfrac{2  \cos \theta \cosh^{2}(gz \theta^{'}_{\lambda})  }{4 \cosh^{2}(gz \theta^{'}_{\lambda}) }\nonumber\\
	&=&\dfrac{2+ \theta_{\lambda }^{'2} }{4}
\end{eqnarray}
Thus, $P_{\text{coin}}$ approaches a constant value, $ \frac{4}{6+ \theta_{\lambda }^{'2}}$, for a specific $\theta_{\lambda }^{'} = 2\sqrt{(\frac{\gamma}{2g})^{2}-1 } $.

\end{document}